# Advanced Deep Learning Methodologies for Skin Cancer Classification in Prodromal Stages

Muhammad Ali Farooq[1], Asma Khatoon[1], Viktor Varkarakis[1], Peter Corcoran[1]

[1] National University of Ireland (NUIG) Galway H91CF50, IRELAND
{m.farooq3, a.khatoon1, v.varkarakis1, peter.corcoran}
@nuigalway.ie

**Abstract.** Technology-assisted platforms provide reliable solutions in almost every field these days. One such important application in the medical field is the skin cancer classification in preliminary stages that need sensitive and precise data analysis. For the proposed study the Kaggle skin cancer dataset is utilized. The proposed study consists of two main phases. In the first phase, the images are preprocessed to remove the clutters thus producing a refined version of training images. To achieve that, a sharpening filter is applied followed by a hair removal algorithm. Different image quality measurement metrics including Peak Signal to Noise (PSNR), Mean Square Error (MSE), Maximum Absolute Squared Deviation (MXERR) and Energy Ratio/ Ratio of Squared Norms (L2RAT) are used to compare the overall image quality before and after applying preprocessing operations. The results from the aforementioned image quality metrics prove that image quality is not compromised however it is upgraded by applying the preprocessing operations. The second phase of the proposed research work incorporates deep learning methodologies that play an imperative role in accurate, precise and robust classification of the lesion mole. This has been reflected by using two state of the art deep learning models: Inception-v3 and MobileNet. The experimental results demonstrate notable improvement in train and validation accuracy by using the refined version of images of both the networks, however, the Inception-v3 network was able to achieve better validation accuracy thus it was finally selected to evaluate it on test data. The final test accuracy using state of art Inception-v3 network was 86%.

**Keywords:** Melanoma, CNN, DNN, Dermoscopy, Inception-v3, MobileNet,

## 1 Introduction

Cancer nowadays is one of the greatest growing groups of diseases throughout the world, among which skin cancer is most common of them. According to stats and figures, the annual rate of skin cancer is increasing at an alarming rate each year [1]. The modern medical science and treatment procedures prove that if skin cancer is detected in its initial phase then it is treatable by using appropriate medical measures which includes laser surgery or removing that part of the skin which ultimately could save a patient's life. Skin cancer has two main stages which include malignancy and melanoma among which melanoma is fatal and comes with the highest risk. In most cases, malignant mole is clearly visible on the patient's skin which is often identified by the patients themselves.

Dermoscopic diagnosis refers to a non-invasive skin imaging method, which has become a core tool in the diagnosis of melanoma and other pigmented skin lesions. However, performing dermoscopy using conventional methods may lower down the diagnostic accuracy which can lead to more chances of errors. These errors are generally caused by the complexity of lesion structures and the subjectivity of visual interpretations [18].

Computer-Aided Diagnosis (CAD) system is a type of digitized platform based on advanced computer vision, deep learning, and pattern recognition techniques for skin cancer classification. For the proposed study we have designed a CAD system for skin cancer classification by utilizing advanced deep neural networks. The system consists of the following steps: Firstly, a preprocessing of the digital images which includes removing clutter such as hair from that part of the skin where the pigmented mole is present and applying a sharpening filter to make that area more clear and visible thus minimizing the chances of error. The next essential step includes the feature extraction and classification process to extract the results for the cases under consideration by utilizing deep learning techniques. Section 2 presents the background and related study and highlights the medical aspects regarding skin cancer. Section 3 describes the detailed methodology of the proposed system whereas Section 4 presents the implementation and experimental results of the proposed study. Section 5 draws the overall conclusion of the paper.

## 2    Background/ Related Work

The human skin is the largest organ of the overall human body. It covers all other organs of the body. It guards the entire body from microbes, bacterium, ultraviolet radiation, helps to regulate body temperature and permits the sensations of touch, heat, and cold [2].

### 2.1    Skin Moles and Skin Cancer

Mole or nevus on human skin can be described as a dark, erected spot comprised of skin cells that are grown in a group rather than individually. These cells are generally known as melanocytes which are responsible for producing melanin, the pigment color in our skin. The main reason behind mole development on human skin is predominantly because of direct sun exposure and any kind of extreme injury. The fair skin population has a greater ratio of skin moles due to the lower quantity of melanin (natural pigments) in their skins [3]. There are three different kinds of skin malignant growth, which include Basal Cell Carcinoma (BCC), Squamous Cell Carcinoma (SCC), and Melanoma. Malignancy is a description of the "stage" of cancer. These malignant growths are critical however, Melanoma comes with the highest risk level and it is discovered more frequently in individuals maturing under 50 years for men and over 50 years for women [4].

## 2.2 Related Work/ Previous Studies

The study proposed by Simon Kalouche utilizes [5] computer vision-based deep learning methods to detect skin cancer and more specifically melanoma. Their dataset was trained on 3 different learning models including a logistic regression model, fine-tuned VGG-16 and multi-layer perceptron deep neural network to achieve a significant amount of classification accuracy. Their results show that their algorithm ability to segment moles and classify skin lesions is 70% to 78%. Md Zahangir et al [6] presented the Inception Recurrent Residual Convolutional Neural Network (IRRCNN) method for breast cancer classification on BreakHis and some other publicly available datasets. They compared their experimental results against existing machine learning techniques in terms of patch-based, image-based, patient-level and image-level classification. Their IRRCNN models provide efficient classification in terms of Area Under the Curve (AUC), global accuracy and the ROC curve. Andre Esteva et al [7] demonstrated classification of skin cancer using a single CNN, which they have achieved by end to end training using image data which is based on disease and pixel labels. In their work, they utilized a large dataset of clinical images that consist of several diseases.

## 3 Methodology

In the proposed study, an efficient skin cancer diagnosis system has been implemented for precise classification between malignant melanoma and benign cases. The complete algorithm consists of several steps starting from the input phase of applying image preprocessing ranging to the analysis of the case under consideration in the form of the probability of lesion Malignancy. Fig. 1 shows the complete workflow of the proposed algorithm.

### 3.1 Image Preprocessing

For the proposed study, the Kaggle skin cancer dataset [8] consisting of processed skin cancer images of ISIC Archive [9] has been utilized. The dataset has a total of 2637 training images and 660 testing images with a resolution of 224 x 224. It is consists of two main classes which include melanoma and benign cases. For image preprocessing two major operations have been applied which includes an initial sharpening filter followed by hair removal filter using dull razor software [10]. These were selected in order to remove the clutter. The results of the image preprocessing operations on two random sample cases are shown in Fig. 2.

It is noteworthy that image quality is refined after applying the image preprocessing operations. This is shown in Section 4 of the paper where the results from four different image quality metrics Peak Signal to Noise Ratio (PSNR), Mean Squared Error (MSE), Maximum Absolute Squared Deviation (MXERR) (MXERR) and Ratio of Squared Norms (L2RAT) on both ground truth images, and preprocessed images are presented.

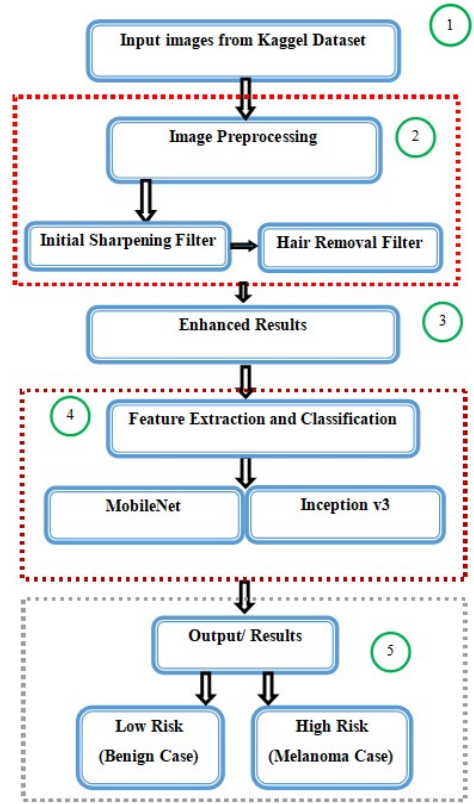

**Fig 1.** Workflow diagram of the proposed method

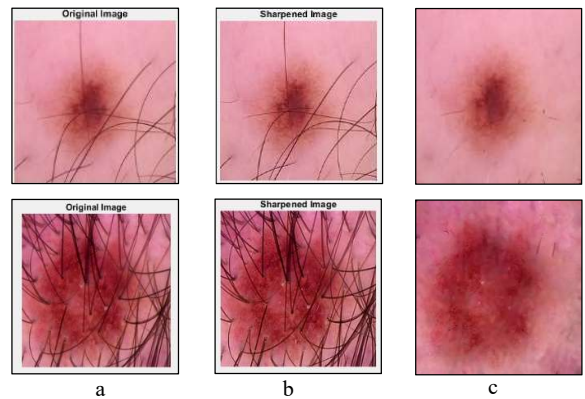

**Fig. 2.** Image preprocessing operations a) original image, b) initial Matlab sharpening filter, c) hair removal using dull razor software [10]

### 3.2 Feature Extraction and Classification

In the next step, the processed images are fed to state-of-the-art deep neural networks in order to perform the feature extraction and classification steps. In this work, the Inception-v3 and MobileNet deep learning architectures are utilized. These architectures play a vital role by extracting feature values from raw pixel images. The Inception-v3 has state of the art performance in the classification task. It is made up of 48 layers stacked on top of each other [11]. The Inception-v3 model was initially trained using 1.2 million images from Imagenet [12] of 1000 different categories. These pre-trained layers have a strong generalization power and they are able to find and summarize information that will help to classify most of the images from the real-world environment. For the proposed study we have utilized this network for our custom classification task by retraining the final layer of the network thus updating and finetuning the softmax layer, by applying the method of transfer learning. This was preferred as the amount of data available for this task is limited and training the Inception-v3 from the beginning would require a lot of time and computational resources. Therefore, by fine-tuning the inception v3 model, we take advantage of its powerful pre-trained layers and thus being able to provide satisfying accuracy results even with a limited amount of data. MobileNet is one of the other finest deep learning architectures proposed by Howard et al. 2017 [13] specifically designed for mobile and embedded vision applications. MobileNet is counted as a lightweight deep learning architecture. It uses depth-wise separable convolutions that means it performs a single convolution on each color channel rather than combining all three and flattening it. This has the effect of filtering the input channels. For our experiments, the networks were trained with two different types of data. The networks were trained with the original images and also with the images after applying the preprocessing operations to them. The training and validation accuracy were examined in order to study the effect of the training on the networks with the two different types of data. Finally, the accuracy on the test set is calculated in order to evaluate the overall performance of the classifiers

## 4 Implementation and Experimental Result

The overall algorithm was implemented using Matlab R2018a for computing image quality metrics and TensorFlow [14] for training the classifiers. The system was trained and tested on a Core I7 sixth-generation machine equipped with NVIDIA RTX 2080 Graphical Processing Unit (GPU) having 8GB of dedicated graphic memory. The first part of the experimental results displays the image quality metrics measured for both benign and malignant melanoma cases before and after applying the image preprocessing operations. It is displayed in Table 1.

**Table 1.** Image Quality Metrics

| Image | PSNR | MSE | MAXERR | L2RAT | Dimension |
|---|---|---|---|---|---|
| 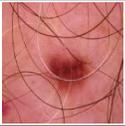 | 19.4205 | 743.0656 | 99 | 0.9657 | 224 x 224 |
| 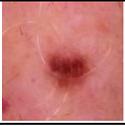 | 21.5481 | 655.2738 | 99 | 0.9801 | 224 x 224 |
| 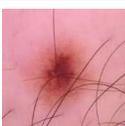 | 22.1285 | 398.3229 | 99 | 0.9868 | 224 x 224 |
| 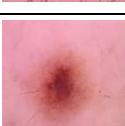 | 23.2953 | 304.4737 | 99 | 0.9902 | 224 x 224 |
| 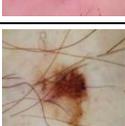 | 22.4291 | 371.6785 | 99 | 0.9852 | 224 x 224 |
| 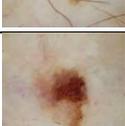 | 24.0840 | 329.9128 | 99 | 0.9903 | 224 x 224 |
| 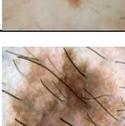 | 18.6732 | 882.5930 | 99 | 0.9681 | 224 x 224 |
| 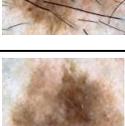 | 19.3975 | 847.0221 | 99 | 0.9747 | 224 x 224 |

The experimental results show clearly that image quality is not comprised however it is upgraded which is evident from high PSNR values and other metrics after applying image preprocessing operations especially the hair removal filter. The image quality metrics were carried out on more than fifty images and the same observations were measured. The second part of the experiments includes the training of the classifiers using the two state of the art deep learning networks i.e. Inception-v3 and MobileNet. For Inception-V3 the data was resized to 299 x 299 since the network has an image input size of 299 by 299. The classifiers were trained on both sets of images i.e. original (ground truth) images and images after applying the preprocessing operations to them. Both the networks were trained using the same hyperparameters. The learning rate was set to 0.005 with a batch size of 32 and total iterations were set to 5000. The training data was split in the ratio of 75% and 25% for training and validations images respectively. Fig. 3 and Fig. 4 display the training and validation accuracy graphs along with the error rate (cross-entropy) graph of MobileNet and Inception-v3 networks.

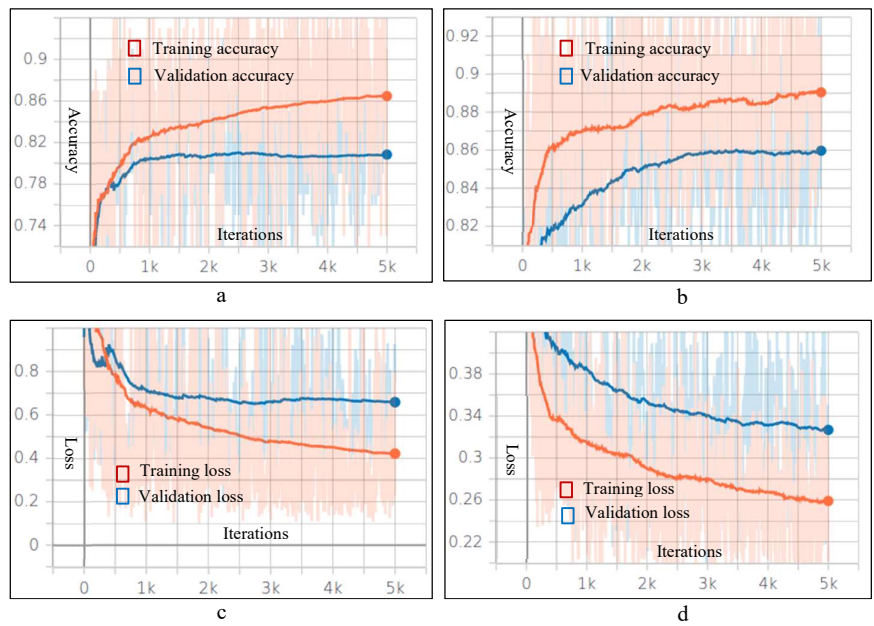

**Fig. 3.** Accuracy and loss graph of MobileNet network a), training and validation accuracy before applying image preprocessing operations b), training and validation accuracy after applying image preprocessing operations c), training and validation loss before applying image preprocessing operations and d) training and validation loss after applying image preprocessing operations

The accuracy graphs in Fig. 3 show that training and validation accuracy before applying the image preprocessing was 86% and 79.8% and it was increased to 89% and 85.9% by using a refined version of images obtained after applying the image preprocessing operations. Similarly, the validation error rate was also decreased from 61% to 32% by using the refined version of images.

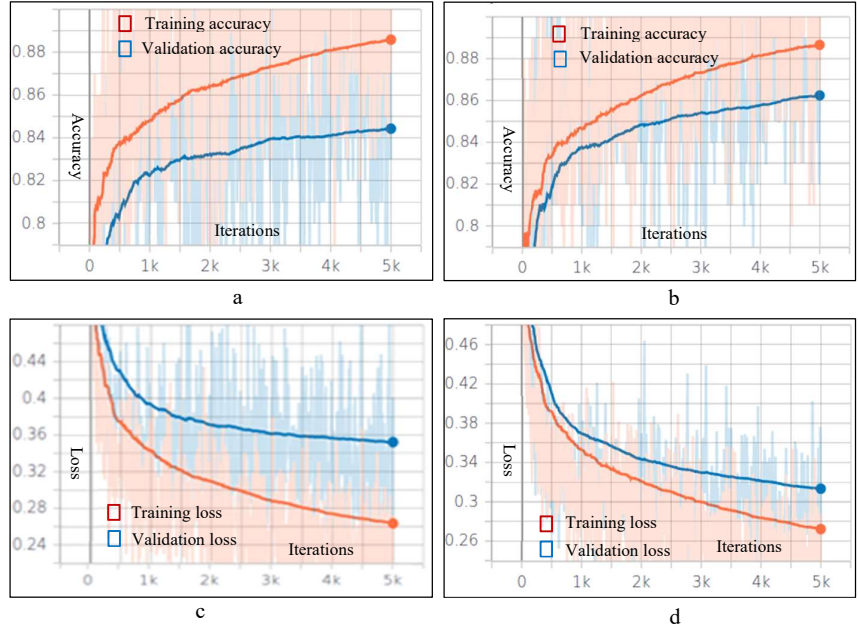

**Fig. 4.** Accuracy and loss graph of Inception-v3 network a), training and validation accuracy before applying image preprocessing operations b), training and validation accuracy after applying image preprocessing operations c), training and validation loss before applying image preprocessing operations and d) training and validation loss after applying image preprocessing operations

The accuracy graphs in Fig. 4 show that training and validation accuracy before applying the image preprocessing was 88.3% and 84.2%. By using the refined version of images training accuracy tends to remain the same thought the validation accuracy was increased to 86.1%. Similarly, the validation error rate was also decreased from 36% to 32.3% by using the refined version of images.

Overall, in both networks, significant improvements were measured after using the refined version of images. The experimental results show that the Inception-v3 network was able to achieve better validation accuracy using a refined version of training data i.e. 86.1 % thus we will be using the Inception-v3 network for evaluating it on the test data. For evaluating the classifiers on the test data, we have picked numerous cases from the test set from both classes, benign and malignant melanoma among which visually complex and challenging test cases were selected for the proposed research work. It is pertinent to mention that the network was tested using the original images (unrefined version) to test the overall effectiveness of the classifier. Fig. 5 shows some of the results predicted correctly on test images. Table 2 illustrates the complete results on visually complex test cases selected for the proposed study which will be further used to evaluate overall testing accuracy, sensitivity (true positive rate), specificity (true negative rate) and precision metrics. The rows highlighted with red color indicates the misclassified test cases when compared with ground truth results.

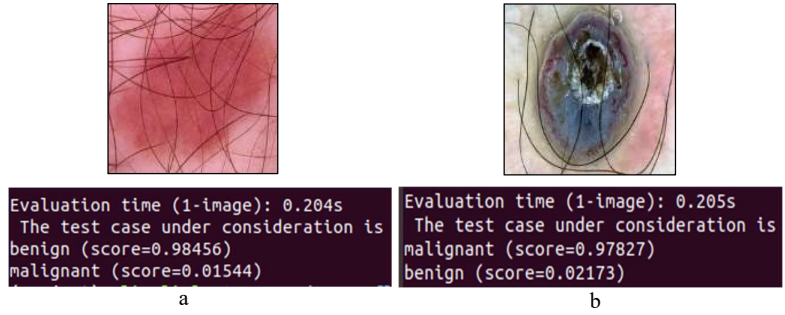

**Fig. 5.** Test case results on two random cases using Inception-v3 network a) case 4 – (benign = Low risk = 98.4% confidence level), b) Case 16 – (malignant melanoma = high risk = 97.8% confidence level).

**Table 2.** Individual Test Case Results

| Test Case | Predicted results using Inception-v3 Network trained on original images | Predicted results using Inception-v3 Network trained on processed images | Ground truth Results |
|---|---|---|---|
| 1 | Benign – Low risk – 97.8% | Benign – Low risk – 84.6% | Low risk |
| 2 | Malignant–High risk – 91.8% | Malignant – High risk – 89.1% | High risk |
| 3 | Benign – Low risk – 98.4% | Benign – Low risk – 96.9% | Low risk |
| 4 | Benign – Low risk – 98.4% | Benign – Low risk – 98.4% | Low risk |
| 5 | Malignant – High risk – 96.4% | Malignant – High risk – 95.7% | Low risk |
| 6 | Malignant – High risk – 98.7% | Malignant – High risk – 96.2% | High risk |
| 7 | Malignant – High risk – 98.8% | Malignant – High risk – 97.8% | High risk |
| 8 | Malignant – High risk – 99.4% | Malignant – High risk – 99.3% | High risk |
| 9 | Benign – Low risk – 71.2% | Benign – Low risk – 60.7% | Low risk |
| 10 | Malignant – High risk – 85.9% | Malignant – High risk – 76.8% | Low risk |
| 11 | Malignant – High risk – 99.5% | Malignant – High risk – 99.3% | High risk |
| 12 | Malignant – High risk – 98.5% | Malignant – High risk – 99.2% | High risk |
| 13 | Malignant – High risk – 70.9% | Malignant – High risk – 87.4% | High risk |
| 14 | Benign– Low risk – 74.8% | Malignant – High risk – 92.3 % | High risk |

| 15 | Malignant – High risk – 97.5 % | Malignant – High risk – 98.7 % | High risk |
| 16 | Malignant – High risk – 99.0 % | Malignant – High risk – 97.8 % | High risk |
| 17 | Benign – Low risk – 96.1% | Benign – Low risk – 97.0 % | Low risk |
| 18 | Benign – Low risk – 86.4% | Benign – Low risk – 86.2 % | Low risk |
| 19 | Benign – Low risk – 99.4% | Benign – Low risk – 99.1 % | Low risk |
| 20 | Malignant –High Risk– 50.2% | Benign – Low risk – 59.0 % | Low risk |
| 21 | Malignant – High risk – 81.9 % | Benign – Low risk – 64.1 % | High risk |

The overall performance of the Inception-v3 network on test data has been evaluated using five quantitative measures: Accuracy, sensitivity, specificity, precision and F1 score [15,19]. These measures are computed using the following forms.

$$Accuracy \ (ACC) = \frac{tp + tn}{tp + tn + fp + fn} \ X \ 100 \qquad (1)$$

$$Sensitivity \ (TPR)/ \ Recall = \frac{tp}{tp + fn} \ X \ 100 \qquad (2)$$

$$Specificity \ (TNR) = \frac{tn}{tn + fp} \ X \ 100 \qquad (3)$$

$$Precesion \ (PPV) = \frac{tp}{tp + fp} \ X \ 100 \qquad (4)$$

$$F1 \ Score \ = 2 \ X \ \frac{Precesion \ X \ Recall}{Precesion + Recall} \ X \ 100 \qquad (5)$$

Where $tp, fp, fn, and \ tn$ refer to true positive, false positive, false negative, and true negative. ACC in (1) means overall testing accuracy, TPR in (2) means true positive rate, TNR in (3) refers to true negative rate while PPV in (4) is an abbreviation for positive prediction value.

Table 3 illustrates the results of all the four quantitative measures: Accuracy, sensitivity, specificity, and precision of the Inception-v3 network before and after using the image preprocessing operations on test data. It can be observed that testing accuracy is increased to 86% by training the classifier using the refined version of images.

**Table 3.** Overall Quantitative Metrics Results on Test Data

| Quantitative measures | Inception-v3 network trained on original images | Inception-v3 Network trained on refined (processed) images |
|---|---|---|
| Accuracy | 81% | 86% |
| Sensitivity | 87.5% | 89% |
| Specificity | 77% | 83% |
| Precision | 70% | 80% |
| F1 Score | 77% | 84% |

## 5. Conclusion and Future work

The main purpose of the proposed study was to improve the overall accuracy level of two state of art deep learning networks which include Inception-v3 and MobileNet by using the refined version of skin cancer images obtained after applying image preprocessing operations. The experiments were conducted using the Kaggle Skin Cancer Dataset by applying initial sharpening filter and hair removal algorithms. Initially, we applied these algorithms as image pre-processing mechanisms to remove the clutters thus producing the refined version of images. Different image quality metrics including Peak Signal to Noise (PSNR), Mean Square Error (MSE), Maximum Absolute Squared Deviation (MXERR) and Energy Ratio/ Ratio of Squared Norms (L2RAT) were used to compare the image quality before and after applying the pre-processing techniques. These metrics prove that image quality was upgraded after applying sharpening filter and hair removal algorithms. In the next phase of experimental results, we have seen substantial improvement in training, validation and test accuracy after applying image pre-processing operation. Thus, we have achieved an overall test accuracy of 86% using state of the art Inception-v3 network by fine-tuning the last layer of the network with a refined version of kaggle skin cancer training dataset.

For future work, more image pre-processing techniques like neural networks based super image algorithms and other such techniques could be used to improve the image quality to a better extent. Moreover, other state of the art deep neural networks such as ResNet-101 [16], Xception [17] could be utilized in order to improve the accuracy levels.